\documentstyle[prl,aps,epsfig,twocolumn,amssymb]{revtex}

\begin{document}
  \columnsep0.1truecm

  \title  {Analytical  and Numerical Flash-Algorithms for Track Fits}
  \author {M. Dima}
  \address{Dept. of Physics, Campus Box 390            \\
           University of Colorado,                     \\
           Boulder, CO-80309                                }
  \maketitle

  \begin{abstract}
 Flash-algorithm track-reconstruction routines with
 speed factors 3000-4000 in excess those of 
 traditional iterative routines are presented.
 The methods were successfully tested in the alignment
 of the Test Beam setup for the ATLAS Pixel Detector
 MCM-D modules, yielding a 
 60 fold increase in alignment resolution
 over iterative routines, for the same
 amount of alocated CPU time.
\end{abstract}


\narrowtext

\section{Introduction}

   In many particle physics experiments high-precision alignment
 can be performed 
 provided there is sufficient data and enough computer (CPU) time alocated.
 While statistics may suffice for many
 experiments, CPU time 
 can only be decreased through the use of 
 flash reconstruction routines.
 For instance in pixel detectors,
 alignment parameters
 can be tuned iteratively to satisfy good track reconstruction
  in all events. 
 However, reconstruction of all tracks in all events, 
 for each iteration
  in alignment parameter space\footnote{An example of an analytical
  alignment procedure ({\em vs}. an iterative one) is
   illustrated in ~\cite{sld-th} for the
    SLD End-Cap \v{C}erenkov Ring
 Imaging Detector.},
  is CPU exhausting and
 in practice can only be applied to blocks of
   tracks.
  If the track fits themselves are also iterative~\cite{kw},
   double-nesting 
   of iteration loops occurs and CPU times reach 
  on the order of months.
  On the other hand if 
  track fits are semi/analytical,
 CPU times drop 3-4 orders in magnitude 
 and such an approach becomes feasable.
 
 The paper starts with the basic $\chi^2$ fit,
 examines the validity of the standard
 quadratic approximation
 for $\chi^2$ and presents analytical and
 semi-analytical track reconstruction algorithms,
  together with CPU-clocked 
 examples\footnote{The tests were performed on a DEC-ALPHA 878
 machine, with EV 5.6 processor at 433 MHz, 640 MB RAM,
 and running under OSF1 V4.0. Coding was performed in FORTRAN.}.
 The methods were developed gradually from the simple case
 of straight tracks (magnetic field free environment, or
 high momentum tracks), to the more demanding case of helical tracks.
 The Test Beam setup of the ATLAS Pixel Detector~\cite{inn} MCM-D
 modules~\cite{becks} was used as
 prototype example for the applied methods, although the latter
 methods are general and applicable to a number of detectors using tracks.

\section{General Fits}

  The general expression for $\chi^2$ in a fit is 
 the sum of the squared normalised residuals  over the set of experimental
 points:
 \begin{equation}
 \chi^2 \, \buildrel \rm def \over =\, \sum^{N}_{j = 1}
   \bigg(\frac{\vec{ \, \Delta}_j}{\sigma_{j_{(\Delta)}}}\bigg)^2
 \end{equation}
  where the error of the $j^{th}$-residual
 $\sigma_{j_{(\Delta)}}$ is related to the direction of the residual itself,
 $\vec{ \, \Delta}_j$ through the covariance matrix ${\mathbf \sigma}_j^2$ :
 \begin{equation}
  \sigma^2_{j_{(\Delta)}} = 
  \frac{\vec{ \, \Delta}_j \cdot {\mathbf \sigma}_j^2
 \vec{ \, \Delta}_j }{\big(\vec{ \, \Delta}_j\big)^2}
 \label{eq:chi-gen-2}
 \end{equation}
 leading to:
 \begin{equation}
 \chi^2_{exact} =  \sum^{N}_{j = 1} \frac{\big(\vec{ \, \Delta}_j \big)^2}
 {\vec{ \, \Delta}_j \cdot {\mathbf \sigma}_j^2
 \vec{ \, \Delta}_j } 
 \label{eq:chi-gen-3}
 \end{equation}

  More often however, a 
 quadratic\footnote{$\vec{ \, \Delta}_j \cdot {\mathbf \sigma}_j^{-2}
 \vec{ \, \Delta}_j$ =
 $\Delta^2_x/\sigma^2_x +
  \Delta^2_y/\sigma^2_y +
  \Delta^2_z/\sigma^2_z$, for $\sigma_j^2$ assumed diagonal.}
 approximation of
 $\chi^2_{exact}$ is used:
 \begin{equation}
 \chi^2_{approx} \, \buildrel \rm def \over =\,
 \sum^{N}_{j = 1}    \vec{ \, \Delta}_j \cdot
   {\mathbf \sigma}_j^{-2}
  \vec{ \, \Delta}_j 
 \label{eq:chi-gen-4}
 \end{equation}
 with the equivalent residual error:
 \begin{equation}
 \sigma^2_{j_{(\Delta)}}  =
 \frac {  \big( \vec{ \, \Delta}_j \big)^2
   }{  \vec{ \, \Delta}_j \cdot
   {\mathbf \sigma}_j^{-2}
    \vec{ \, \Delta}_j}
 \label{eq:chi-gen-5}
 \end{equation}
 If a track impacts a point's error ellipsoid
 at $n = tg(\theta_{incid}$) with respect
 to one of the principal axes, then approximating
  $| \Delta_j \rangle$ 
 as perpendicular to the trajectory, the two $\sigma$'s 
 can be written as:
 \begin{eqnarray}
  \sigma^2_{exact} \,  &\simeq& \, \frac{\sigma^2_{xy} + n^2 \sigma^2_{z}}{1 +
  n^2} \cr \cr
 \sigma^2_{approx} \,  &\simeq&  \, \frac{1 + n^2}{\sigma^{-2}_{xy} + n^2 
 \sigma^{-2}_{z}} 
 \end{eqnarray}
 both reducing to $\sigma^2_{xy}$ or
 $\sigma^2_z$ for tracks impacting along one of the
 principal axes.

 The Test Beam stand for the
 ATLAS Pixel MCM-D modules is a Telescope setup, with
  4 tracking elements 
 (Sirocco strip Detectors) and two slots for
 Pixel Module evaluation.
   The setup was mounted on a marble optical-bench,
    itself placed on a rail that allowed it
     to be   moved into the
    active area of  a 1.4 Tesla Spectrometer Magnet.
 The latter was used to determine the magnetic field influence
 (x-shifts) in an environment comparable to the
 ATLAS Detector.   The Sirocco
 strips were mounted along $\Delta z$ = 1.4 m
 of beam-line with a tolerance of $\pm 0.5$ mm.
  The  strips, 30$\mu$m wide, provided a
 resolution of  4$\mu$m
 in the x- and y-directions, while the Pixels,
 50$\mu$m $\times$ 400$\mu$m, a resolution
 on the order of 14$\mu$m $\times$
    180$\mu$m depending on the cast technology~\cite{becks} of
 the chips.
 The error matrices of the Test Beam stand Telescope points
 had thus associated ellipsoids with
 aspect ratios of 1:125,
 the difference between fitting a 3D-line to these
 $\sigma$'s and one to spherical $\sigma$'s being 0.1 $\mu$m
 in the Telescope's mid-plane.

 The difference between $\chi^2_{exact}$ and
 $\chi^2_{approx}$ depends
 strongly on the impact angle of the track onto
 the individual
  error ellipsoids:
 \begin{equation}
  \frac{\Delta \chi^2}{\chi^2_{exact}} \, = \, 
 \frac {\big( \sigma_{xy}/\sigma_z \big)^2 +
        \big( \sigma_z/\sigma_{xy} \big)^2 - 2}{\big( 
   n + 1/n \big)^2}
 \label{eq:chi-gen-6}
 \end{equation}
 For the Telescope setup the track's impact angle was on the average
 0.15 mrad, with a corresponding $\Delta \chi^2/\chi^2_{exact}$
 on the order of 0.02\%. 
 When the tracks impact however at an arbitrary angle,
  $\Delta \chi^2/\chi^2_{exact}$ 
 can reach as high as 3000 for
 the current $\sigma_{z}/\sigma_{xy}$ ratio,
 even if $\Delta \sigma^2 / \sigma^2_{exact}$ is on the order
 of unity. 

  For tracks impacting all points at a constant angle 
  (``stiff"-tracks),  $\Delta \chi^2 $
 is constant along the trajectory
 and the two methods   yield identical results.
 If the track is measured however piecewise
 in two different sub-systems,
 or it is composed of an ensemble of points
 with different $\sigma$'s (different types of detectors),
 then even for straight 
 tracks the two solutions differ.
 For tracks bending in magnetic field,
 the track's impact angle changes continuously
 along the track, $\Delta \chi^2$ following as:
 \begin{equation}
    d\bigg(\frac{\Delta \chi^2 }{ \chi^2_{exact}}\bigg) \bigg / 
   \frac{\Delta \chi^2 }{ \chi^2_{exact}} =  
  \frac{dn}{n}  \, \cdot 2 \, \, \frac{1-n^2}{1+n^2}  
 \label{eq:chi-gen-7} 
 \end{equation}
 Within the 1.4 m of the Telescope, the 180 GeV/c
  tracks used bend in the B = 1.4 T magnetic field 
 equivalently to $\Delta \chi^2/\chi^2_{exact}$ $\simeq$
 0.02\% up front and 
 16\% downstream, the approximative method
 pulling the fit increasingly tighter
 towards the end -
 on the order of 0.5 $\mu$m per point.
  The
 effect is
 evidently insignificant, both in the Telescope setup,
 as well as in the 
 real B-physics context of ATLAS, meaning 2 Tesla magnetic
 field, tracks of momentum greater than 1 GeV/c, and a
 measured track length of approximately 0.14 m.
 Over this span the expected point to point change in $\chi^2$
 is less than  4 \% per \%-$\Delta \chi^2$.

 \section{Line Fits}

  In most cases it is possible to interchange
 the non-linear expression (\ref{eq:chi-gen-3})
 with its quadratic 
 approximation (\ref{eq:chi-gen-4}), allowing
 analytical solutions to be given for ``stiff-tracks"
 {\em i.e.} - 
   particle out
 of magnetic field,
 weak field with respect to track momentum, or
 distance travelled small with respect to
 existing resolution.

 The simplest fit is for
  $\sigma^2_j = \sigma^2 \cdot {\mathbf 1} = const.$ :
\begin{equation}
\chi^2 = \sum_{j = 1}^N \big( \vec{\, \Delta}_j \big)^2 = min.
\label{eq:lin-fit-1}
\end{equation}
Parametrising the tracks as:
\begin{equation}
 \vec{r} = \vec{r}_0 + \lambda \vec{n}
\label{eq:lin-fit-2}
\end{equation}
with $\vec{n}^2 = 1$ and $\vec{r}_0 \cdot \vec{n} = 0$,
 equation (\ref{eq:lin-fit-1}) becomes:
\begin{equation}
\chi^2 = \sum_{j = 1}^N \vec{\delta}_j^2 = min.
\label{eq:lin-fit-3}
\end{equation}   
where $\vec{\delta}_j = \vec{r}_j - \vec{r}_0 - \lambda_j \vec{n}$.
The minimum condition implies locally
 $\lambda_j = \vec{r}_j \cdot \vec{n}$
and globally: 
\begin{eqnarray}
  \vec{r}_0 &=& ({\mathbf 1} - \vec{n}\vec{n})\, \langle \vec{r} \, \rangle
 \cr 
{\mathbf M}\,  \vec{n} &=& \mu_0 \, \vec{n} 
\label{eq:lin-fit-5}
\end{eqnarray}
 where $\langle \vec{r} \, \rangle$ denotes average over 
 measured
 points, and ${\mathbf M} = \langle \vec{r} \, \vec{r} \, \rangle -
 \langle \vec{r} \, \rangle \langle \vec{r} \, \rangle$ the spread
 ellipsoid of points around $\langle \vec{r} \, \rangle$.
 The 3 eigen-values
 of ${\mathbf M}$
 represent the length of the track ($\mu_0$),
 and the two
 transversal variances to the line-fit.

\begin{figure}[t]
\hspace{-0.4cm}
\vspace{-2.2cm}
  \begin{flushleft}
        \mbox{
        \epsfig{file=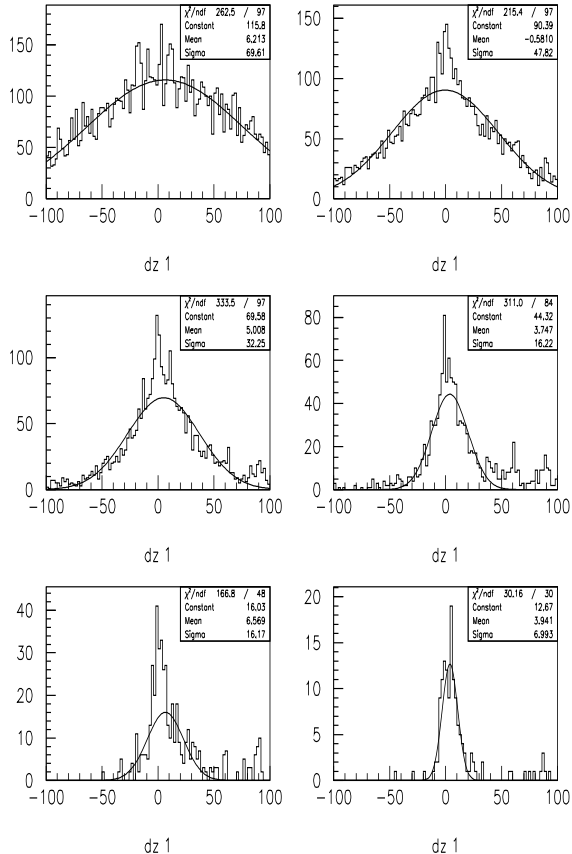,width=8.5cm,height=12.9cm}
             }
  \end{flushleft}
      \caption{Translational alignment parameter ($\Delta z$) histograms
            for blocks of 2, 4, 8, 20, 40 and 80 tracks, for Sirocco Plane 1.
            The effective number of track-fits
            in each histogram is $\simeq$ 3 million.
     Although in the Telescope setup
 the alignment lever arm for $\Delta z$ is very low
 (0.15 $\mu$m/mm, with a $\sigma$ = 4 $\mu$m
 detector resolution), the plots show a dramatic improvement
 in $\Delta z$ resolution
 with
 increasing number of tracks ($N_{block}$) per
 iterative alignment loop. The electronics noise is cut-down by a
 factor of $1/\sqrt{N}_{block}$.
 The approach demands however extensive CPU power,
 analytical solutions being needed in order to keep the problem
  within the capabilities of existing resources. All figures are in $mm$.}
  \label{fig:d-z}
\end{figure}

  For $\sigma^2_j = diag(\sigma^2_x,
 \sigma^2_y, \sigma^2_z) = const.$ equation
 (\ref{eq:lin-fit-3}) holds again, however 
 in normalised form, with 
 $\vec{r}_i \rightarrow \sigma^{-1}  \vec{r}_i$,
 $\vec{r}_0 \rightarrow \sigma^{-1}  \vec{r}_0$ and 
 $\vec{n} \rightarrow \sigma^{-1}  \vec{n}$.
 Adapted to the geometry of the Telescope
  this solution has been clocked to
 0.033 $\mu$s/2D-fit 
 and 2.6 $\mu$s/3D-fit. These times are half or less for
 current 1000-1500 MHz clock machines.
 CPU-wise this allows the alignment to be
 performed using blocks of tracks. The intuitive
 picture of this procedure is aligning two incomplete
 3D images: where there is a track, there is a pixel
 on the image.
  The more tracks in an alignment loop, the more pixels
 per image, and the better the alignment.
 This idealistic picture is cut cold however, by either  
 lack of data, or of CPU power. In most of the contemporary
 HEP experiments the limiting factor is CPU power,
 as the reconstruction of the objects
 used in the alignment (in this case tracks) can be quite
 CPU costly. Eliminating this problem with flash-algorithms
 opens the way to high accuracy alignment.
   The increase in
 alignment resolution is proportional to 
 $\sqrt{N}_{block}$, and the 
  CPU cost to $N_{block}$. Holding CPU time fixed,
 any speed increase can be equivalenced to a resolution
 increase. Thus a factor of 4000 in speed, would be
 equivalent to a factor of 63 in resolution. To exemplify this,
 consider the Test Beam setup, where the tracks impact
 the  Sirocco planes at almost normal incidence.
 At precisely normal incidence any z-misalignment would
 be un-noticed. The tracks   do impact however at an
 angle on the order of 0.15 mrad, giving an alignment
 ``lever arm" for
 $\Delta z$ misalignments 
 on the order of
  0.15 $\mu$m/mm. With the existing Sirocco resolution
 ($\simeq$ 4 $\mu$m) the resolving power (per track) for $\Delta z$
 would be on the order of 27 mm. Increasing the number
 of tracks per one $\Delta z$ misalignment loop
 the resolution improves dramatically, as illustrated in 
 figure \ref{fig:d-z}.

 Any multi-tile pixel detector
  benefiting from
 this high precision alignment method
 would better perform in
 physics involving the resolution of vertices. With the increasing
 energy frontier (LHC, TESLA/NLC) particle ID techniques
 are replaced with other means of signal identification.
 Signal ``concentrators" that can serve this purpose
  are the B-mesons. Their identification
  depends crucially on vertexing resolution\footnote{For example
 selecting B-events with
 high purities and efficiency can be achieved by applying
 a minimally missing $\vec{p}_\perp$ correction to
 the m$_\pi$ evaluated vertex mass~\cite{cern-b,sld-b}. In the ATLAS
 context $\vec{p}_\perp$ would most likely be referenced to
 the axis of the jet containing the B sub-jet.
  The method depends crucially
 on the vertexing accuracy, yielding for instance at SLD~\cite{sld-b} 
 B-events with
 sample purities on the order of 91-99 \% and 65-20\% corresponding
 efficiencies.}.

  An
  improvement to the example considered would be 
   to include 
 in the track fit also
 the Pixel Demonstrator
 points.
 This would mean fitting to points with different
 error ellipsoids and the imposibility
 of ``absorbing" all
 $\sigma$'s
 into $\vec{r}_0$ and $\vec{n}$ in an unique way, as done previously.
  For such  
 $\sigma^2_i \ne \sigma^2_j$
 cases, the solution is
 given by a set of self-consistent equations:
  \begin{eqnarray}
  \lambda_j &=& \frac{\vec{n} \cdot  {\mathbf \sigma}^{-2}_j
 (\vec{r}_j - \vec{r}_0)}{\vec{n}
  \cdot {\mathbf \sigma}^{-2}_j
 \vec{n}}
 \cr \cr
 \vec{r}_0 &=& \langle \vec{r} \, \rangle -
  \langle \lambda \rangle \vec{n} \cr \cr
    \vec{n} &=&  \frac{\langle \lambda \vec{r} \, \rangle - \langle \lambda
 \rangle \langle \vec{r} \, \rangle}{\langle \lambda^2 \rangle - 
 \langle \lambda \rangle^2}
  \end{eqnarray}
  solvable semi-analytically in approximately 3 iterations, starting
 from the previous analytically exact solution.

\begin{figure}[t]
\hspace{-0.4cm}
\vspace{-2.2cm}
  \begin{flushleft}
        \mbox{
        \epsfig{file=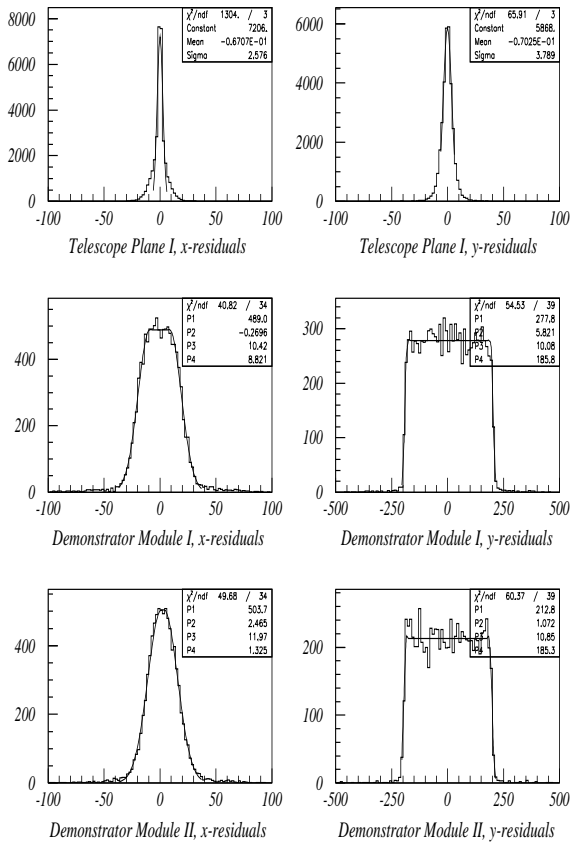,width=8.5cm,height=12.9cm}
             }
  \end{flushleft}
      \caption{Final alignment residuals for the Telescope's Sirocco
  Plane-1 (top),
    $\sigma_x$ $\simeq$ 3 $\mu$m, $\sigma_y$ $\simeq$ 4 $\mu$m,
 and for two Pixel Detectors under evaluation, cast in different 
 technologies (middle and bottom).
  The equal \emph{side}-resolution
 of the Pixels (fit parameter P3, $\sigma_{side} \simeq 11 \mu$m,
 essentially a measure
 of charge division resolution between pixels)
  in the $x$ and $y$-directions
 gives confidence that the alignment conducts to physically meaningful
 results.
 Fit parameter P4 
  gives the single-hit resolution of the pixel ``hat"
 ($\sigma_{hat} = 1 \, ... \, 10 \mu$m, basically the pixel's geometrical
 resolution minus the {\em side}-resolution).
 The center ``thinning" of the Sirocco $x$-residuals
 (from a gaussian) by $\simeq$ 1 $\mu$m was due to
 the residual magnetic field of the Spectrometer Magnet, as 
  Lorentz drift of the charge carriers in the strips.
 All figures are in $\mu m$. }
  \label{fig:fin}
\end{figure}

 The approach used in this paper was to
 tune the alignment\footnote{The alignment consisted of two separate stages,
  firstly out of magnetic field, and secondly in magnetic
   field. In both cases the translational and rotational degrees of
   freedom were firstly determined for the 4 Sirocco planes. The
   Pixel Modules were aligned with
  respect to the Telescope's Sirocco planes.
    The alignment in
  magnetic field required an external marker, which was the
  setup's trigger: a PIN Diode. Its change in image with magnetic field,
   on a ``focal" plane,
   determined the absolute shifts in the tracking elements
   of the Telescope setup (the 4 Sirocco planes),
    and by reference, those of the Pixel
   Modules under evaluation.}
    by looping over the reconstruction of 2000 tracks,
 the alignment residuals of the Telescope Sirocco plane-1 being shown in
 figure \ref{fig:fin} (top). The Pixel Detectors under evaluation
 were 
 to first order
 aligned analytically (exact solution), and subsequently
 tuned in a fashion 
 similar to that of the Sirocco planes,
 in order to compensate for the
 effect of the low $y$-resolution on the better $x$-resolution.
 The 
 difference between the Pixel demonstrator measured and Telescope
 predicted positions of the track hits is shown in
 figure \ref{fig:fin} (middle and bottom)
 for a set of two Pixel Detectors
 under test.
 The identical \emph{side}-resolution of the pixels in the
 $x$ and $y$-directions
 (fit parameter P3) - expected from uniform technology on
 the chip - gives credit that the alignment
 conducted to
 physically tangible results.

\section{Helix Fits}

  Track fits over small arcs of helices are very sensitive to the
 fluctuations of the experimental points. The radius of curvature
 (giving the momentum, or conversely the magnetic field) has consequently
  a large
 error, due to the points moving within their error bars.
 To reduce the errors of such ``weak" parameters,
 fits to collections of tracks - experiencing the same conditions
 (magnetic field in this case) are used.
 Iterative fit routines consum
 however between 90000-300000 $\mu$s/3D helix-fit,
 the previously illustrated method (figure \ref{fig:d-z})
 being inapplicable in this case.
  On the other hand, although desired,
 flash semi-analytical algorithms for helix-fits are
 non-trivial to derive.
  Most of contemporary
 High Energy Physics experiments however involve large energies
 and   
 the helical tracks span typically less
 than 1$^o$ of an arc (0.2$^o$ in the case of the Telescope).
 It is possible
 then in such cases
 to perform 3D-helix fits semi-analytically, by perturbatively curving
 a line-fit into a helix-fit.

The helix equations are:
 \begin{eqnarray}
  d_t \, \vec{p} &=& \frac{e c^2}{E} \, \vec{p} \times \vec{B} \cr \cr
  d_t \, \vec{r} &=& c^2 \, \vec{p}/E
 \label{eq:motion}
 \end{eqnarray}
 where $E$ is the particle's energy, 
 $\vec{p}$ its
 momentum and $\vec{B}$ the magnetic field. 
 The solution to equations (\ref{eq:motion}) is:
 \begin{equation}
 \vec{r} = \vec{r}_0 + \lambda \vec{n} +
 \frac{\lambda^2}{2 !} f\big(\frac{\lambda}{R}\big)
 \, {\mathbf F} \, \vec{n} +
 \frac{\lambda^3}{3 !} g\big(\frac{\lambda}{R}\big)
 \, {\mathbf G} \, \vec{n}
 \label{eq:sol-hx}
 \end{equation}
 where $\lambda = \vec{v}_0 t = \omega R t$ is a ``linear" distance
 travelled by the particle,
 $\vec{\omega} = |e|c^2 \vec{B}/E$ the helical rotation pulsation,
 $\vec{n} = \vec{p}_0/p_0$ the direction of engagement of the
 particle onto the magnetic field,
 $\vec{n}_B = e \vec{B} / |e| B$, $R = p_0/|e| B$ 
 a parameter related to the radius of curvature of the helix
 $R_{helix} = R \sqrt{1 - (\vec{n} \cdot \vec{n}_B)^2}$,  
 $f(\zeta)$ and $g(\zeta)$ two functions:
 \begin{eqnarray}
 f(\zeta) &=& \frac{2 !}{\zeta^2} (1 - cos \zeta) 
 \, \buildrel \rm {\zeta \rightarrow 0} \over \simeq \, 
 1 \cr \cr
 g(\zeta) &=& \frac{3 !}{\zeta^3} (\zeta - sin \zeta)
  \, \buildrel \rm {\zeta \rightarrow 0} \over \simeq \,
 1  
 \end{eqnarray}
  respectively 
 ${\mathbf F}$ and ${\mathbf G}$ two tensors:
 \begin{eqnarray}
 {\mathbf F} &=& \times \, \vec{C} \cr
 {\mathbf G} &=&  \vec{C} \, \vec{C} - \vec{C}^2 \cdot {\mathbf 1}
 \end{eqnarray}
 that satisfy ${\mathbf F}^\dagger = - {\mathbf F}$,
 ${\mathbf G}^\dagger =  {\mathbf G}$, 
 ${\mathbf F}\, {\mathbf G} = {\mathbf G}\, {\mathbf F} = - \vec{C}^2 {\mathbf F}$,
 ${\mathbf F}^{\, 2} = {\mathbf G}$, and
 ${\mathbf G}^2 = - \vec{C}^2 {\mathbf G}$. The vector
 $\vec{C}$ is $\vec{n}_B / R$.

  It is evident that for $R \rightarrow \infty$ (or
 equivalently $\lambda \rightarrow 0$),
 expression  
 (\ref{eq:sol-hx})
 reduces to the parametrisation of the line 
 (\ref{eq:lin-fit-2})
 used in performing 
 line fits - which
 is the requirement for the perturbative approach.
 In most
 experiments the third order approximation 
 $f(\zeta) \simeq 1$ and $g(\zeta) \simeq 1$
  holds up to the following limiting factors:
 \begin{itemize}
 \item{{\bf geometric} - the arc of helix should not exceed a length
        beyond the approximation validity for
        $f(\zeta)$ and $g(\zeta)$. This is related
        to the demanded
        resolution $\sigma$ and the particle's momentum:
        \begin{equation} p \,
         \ge \frac{\lambda}{16(\sigma /\lambda)^{1/3}} \simeq
         5 \, \, {\rm GeV/c}
        \end{equation}
        where in the above, $\lambda$ and $\sigma$ are expressed in [m]
        and $p$ in [GeV/c]. The value for the momentum is
        for the Telescope setup.}
 \item{{\bf dE/dx} - the loss of energy along the trajectory
                     determines a ``tighter" helix, the deviation:
                     \begin{equation} 
   \sigma = \frac{\lambda^2 E}{2\pi p^2 c^2} \bigg( \frac{dE}{dx} \bigg)
        \end{equation}
 needing to be smaller than 4 times the allowed tolerance in the
 Pixel plane.}
 \item{{\bf multiple scattering} - multiple deviations from
        the direction of flight add up to
         a displacement of:
        \begin{equation} 
   \sigma \simeq 0.6 \lambda \theta_{rms}
         \end{equation}
        where $\lambda$ is expressed in [mm], 
 $\theta_{rms}$~\cite{ref:pdgg} in [mrad]
 and $\sigma$ in [$\mu$m]. This should determine an error
 in the Pixel plane no larger than the allowed tolerance.}
 \end{itemize}
  The semi-analytical helix fit procedure has 3 steps:
 \begin{enumerate}
 \item{Estimation of $\vec{n}$, the engagement direction of the 
 particle onto the magnetic field. This is obtained
 with small CPU demand via a 3D line flash-fit to the
 first 3-4 points of the trajectory. The vector $\vec{n}$ is
 an eigen-vector of ${\mathbf M} = \langle \vec{r} \, \vec{r} \, \rangle
 - \langle \vec{r} \, \rangle \langle \vec{r} \, \rangle$, hence 
 any perturbation $\delta {\mathbf M} = {\mathbf M}_{\, helix} -
 {\mathbf M}_{\, line}$ 
 changes it only to second  
 order, and for numerical purposes $\vec{n}$ can be considered constant.}
 \item{Using the $\vec{n}$ found above, the second order term corrections
 to $\vec{r}_0$ and $\lambda_i$ can be estimated:
 \begin{eqnarray}
  \Delta \lambda_i &=& \frac{\lambda_i}{R} (\vec{r}_i - \vec{r}_0)
        \cdot \vec{n} \times \vec{n}_B \cr
  \Delta \vec{r}_0 &=& \langle \vec{r} \, \rangle - \vec{r}_0 -
      \langle \lambda \rangle \vec{n}
 - \frac{\langle \lambda^2 \rangle}{2 R}
               \vec{n} \times \vec{n}_B 
 \end{eqnarray}
 computations again only modestly CPU demanding.}
 \item{ Introducing the third order term and using 
 the previously corrected 
 $(\vec{n}, \vec{r}_0, \lambda_i)$,
 local and global equations for the parameters can
 be written:
 \begin{eqnarray}
  &(& \vec{n} + \lambda_i {\mathbf F} \, \vec{n} + \frac{\lambda_i^2}{2}
 {\mathbf G} \, \vec{n}) \cdot \vec{\rho}_i = 0 \cr
  &\langle& \, (\lambda \cdot {\mathbf 1} - \frac{\lambda^2}{2}
 {\mathbf F} + \frac{\lambda^3}{6} {\mathbf G}) \vec{\rho} \, \,
 \rangle = 0
 \cr
 &\langle& \, \vec{\rho} \, \, \rangle = 0
 \end{eqnarray}
 where $\vec{\rho}_i$ are the residuals
 of the points to the fitted curve:
 \begin{equation}
  \vec{\rho}_i = - \vec{r}_i +
 \vec{r}_0 + \lambda_i \vec{n} + \frac{\lambda_i^2}{2}{\mathbf F} \, \vec{n} +
 \frac{\lambda_i^3}{6} {\mathbf G} \, \vec{n} 
 \end{equation}
 Expanding to first order, the corresponding corrections 
 $(\Delta \vec{n}, \Delta \vec{r}_0, \Delta \lambda_i)$
 must satisfy:
 \begin{eqnarray}
 &{\mathrm \,}& \alpha_i \Delta \lambda_i +
  \vec{a}_i \cdot \Delta \vec{n} +
  \vec{b}_i \cdot \Delta \vec{r}_0 +
  \beta_i  =  0 \cr
   &\langle& \vec{a} \, \Delta \lambda \rangle +
  \langle \lambda^2 \rangle {\mathbf 1} \cdot \Delta \vec{n} +
  {\mathbf D} \, \Delta \vec{r}_0 + \vec{\delta}  =  0 \cr
   &\langle& \vec{b} \, \Delta \lambda \rangle +
  {\mathbf D}^\dagger \Delta \vec{n} +
  {\mathbf 1} \cdot \Delta \vec{r}_0 + \vec{\sigma}  =  0 
 \label{eq:glb}
 \end{eqnarray}
 where:
 \begin{eqnarray}
 \alpha_i &=& \vec{n}^2 - \vec{r}_i \cdot  {\mathbf F} \, \vec{n} +
              \vec{r}_0 \cdot  {\mathbf F} \, \vec{n} \cr \cr
 \beta_i &=& \vec{r}_0  \cdot  \vec{n} - \vec{r}_i  \cdot  \vec{n} +
              \lambda_i \vec{n}^2 + \lambda_i
              \vec{r}_0 \cdot  {\mathbf F} 
 \, \vec{n} - \cr \cr &{\mathrm \,}& \, \lambda_i 
 \vec{r}_i \cdot  {\mathbf F} \, \vec{n} +  
 \frac{1}{2}  \lambda_i^2 \vec{r}_0   \cdot {\mathbf G} \, \vec{n} -
 \frac{1}{2} \lambda_i^2 \vec{r}_i \cdot   {\mathbf G} \, \vec{n} + \cr \cr
 &{\mathrm \,}& \,
   \frac{1}{6} \lambda_i^3 \vec{n} \cdot  {\mathbf G} \, \vec{n}  \cr \cr  
 \vec{a}_i &=& \vec{r}_0 - \vec{r}_i + 2\lambda_i \vec{n} +
               \lambda_i {\mathbf F} \, \vec{r}_i - 
               \lambda_i {\mathbf F} \, \vec{r}_0 \cr \cr
 \vec{b}_i &=& \vec{n} + \lambda_i {\mathbf F} \, \vec{n} \cr \cr
 \vec{\delta} &=& \langle \lambda \rangle \vec{r}_0 -
             \langle \lambda \vec{r} \, \rangle +
            \langle \lambda^2 \rangle \vec{n}
            + \frac{1}{2} {\mathbf F} \,   \langle
 \lambda^2 \vec{r} \, \rangle - \cr \cr
 &{\mathrm \,}& \,  
              \frac{1}{2} 
\langle \lambda^2 \rangle {\mathbf F} \, \vec{r}_0 +  
              \frac{1}{6} \langle \lambda^3 \rangle 
{\mathbf G} \, \vec{r}_0 -  
              \frac{1}{6} {\mathbf G} \,   \langle 
\lambda^3 \vec{r} \, \rangle +  \cr \cr
 &{\mathrm \,}& \,   
             \frac{1}{12}  \langle \lambda^4 \rangle
 {\mathbf G} \, \vec{n}
              \cr \cr
  \vec{\sigma} &=& \vec{r}_0 - \langle \vec{r} \, \rangle +
          \langle \lambda \rangle \vec{n} +
           \frac{1}{2} \langle \lambda^2 \rangle {\mathbf F}
          \, \vec{n} + \frac{1}{6} \langle \lambda^3 \rangle {\mathbf G}
          \, \vec{n} \cr \cr
 {\mathbf D} &=& \langle \lambda \rangle {\mathbf 1}
   - \frac{1}{2} \langle \lambda^2 \rangle {\mathbf F}  
 \end{eqnarray}  
 By eliminating $\Delta \lambda_i =
   -  (\beta_i +
            \vec{a}_i \Delta
  \vec{n} + \vec{b}_i \Delta \vec{r}_0 ) / \alpha_i$
  equations (\ref{eq:glb}) become:
 \begin{eqnarray}
     {\mathbf M} \, \Delta \vec{n} + {\mathbf N} \, \Delta \vec{r}_0 &=&
  \vec{\tau} \cr \cr
   {\mathbf N}^\dagger \Delta \vec{n} + {\mathbf R} \, \Delta \vec{r}_0 &=&
  \vec{\pi}  
 \end{eqnarray}  
 where:
 \begin{eqnarray}
     {\mathbf M}  &=& 
  \big< \frac{\vec{a} \, \vec{a}}{\alpha}   \big> -
     \langle \lambda^2 \rangle {\mathbf 1} \cr \cr
     {\mathbf N}  &=& 
  \big< \frac{\vec{a} \, \vec{b}}{\alpha}   \big> -
       \langle \lambda  \rangle  {\mathbf 1} 
 + \frac{1}{2}   \langle  \lambda^2 \rangle   {\mathbf F} \cr \cr
     {\mathbf R}  &=& 
  \big< \frac{\vec{b} \, \vec{b}}{\alpha}   \big> -
        {\mathbf 1}  
 \end{eqnarray}  
  and :
 \begin{eqnarray}
     \vec{\tau} &=& \vec{\delta} -   \big<
  \frac{\beta \, \vec{a}}{\alpha}   \big> \cr \cr
     \vec{\pi}  &=&  \vec{\sigma} -   \big<
  \frac{\beta \, \vec{b}}{\alpha}   \big>  
 \end{eqnarray}  
  To zero$^{th}$ order the three ${\mathbf M}$,
 ${\mathbf N}$ and ${\mathbf R}$
 tensors are proportional
 to $({\mathbf 1} - \vec{n}\vec{n})$, the non-invertable
 perpendicular projection to $\vec{n}$, by factors of
 $\langle \lambda^2 \rangle$, $\langle \lambda \rangle$
 and 1.
 Therefore in the numerical approach,
 the inversion is obtained by decomposing the operators into
 a part proportional to $({\mathbf 1} - \vec{n}\vec{n})$
 and a ``remainder":
 \begin{equation}
 {\mathbf R} = ({\mathbf 1} - \vec{n} \vec{n}) \cdot
 (2 tr {\mathbf R} - {\mathbf R}_{\vec{n} \, \vec{n}}
 - {\mathbf R}^\dagger_{\vec{n} \, \vec{n}}) / 4 + ...
 \end{equation}
 The solution $(\Delta \vec{n}, \Delta \vec{r}_0, \Delta \lambda_i)$
 is therefore:
 \begin{eqnarray}
     \Delta \vec{n} &=& ({\mathbf R} {\mathbf N}^{-1} {\mathbf M} -
 {\mathbf N}^\dagger)^{-1} ({\mathbf R} {\mathbf N}^{-1} \vec{\tau} -
 \vec{\pi}) \cr \cr
  \Delta \vec{r}_0 &=& {\mathbf N}^{-1} (\vec{\tau} - {\mathbf M}
 \Delta \vec{n}) \cr \cr
  \Delta \lambda_i &=&
   -  \frac{1}{\alpha_i}(\beta_i +
            \vec{a}_i \cdot \vec{n} + \vec{b}_i \cdot \Delta \vec{r}_0 )  
 \end{eqnarray}  
 All quantities 
  in this section were considered normalised
 - \emph{i.e.} $\sigma^{-1} \vec{r} \rightarrow \vec{r}$,
 although for notation simplicity they were written as the quantities 
 themselves.
 }
 \end{enumerate}

   The CPU demand of the above 3 steps is under 15 $\mu$s. For 
 better precision however, the last step can be repeated twice,
 bringing the 3D-helix fit to 22 $\mu$s. This is
  at least 4000 times faster than any iterative version
 of the fit.

 To complete the fit in all generality,
 $dE/dx$ can be incorporated by
 perturbatively
 bending a pure helical track into a $dE/dx$-helix.
 This would 
 allow in principle to infer the track's mass,
 although with large errors in certain energy ranges.

\begin{figure}[t]
\hspace{0.4cm}
\vspace{-2.0cm}
  \begin{flushleft}
        \mbox{
        \epsfig{file=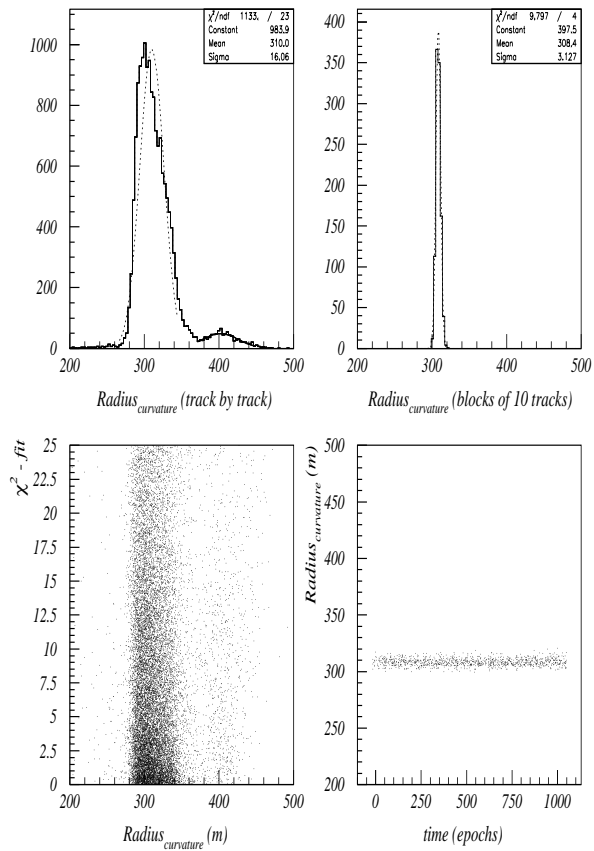,width=8.5cm,height=12.9cm}
             }
  \end{flushleft}
      \caption{Normal-impact equivalent radius of curvature for
 single tracks (top-left) and corresponding
 $\chi^2$ of fit \emph{vs.} radius of curvature (bottom-left).
 The resolution was on the order of 16 m and the average radius of
 curvature 310 m. Using blocks
 of 10 tracks (top-right) the resolution
 improves beyond the statistical factor, 
 to $\simeq$ 3.3 m, the stability of the SPS delivered
 beam being tracked \emph{vs.} time bottom-right. }
  \label{fig:rrr}
\end{figure}

 Using the fit over blocks of 2000 tracks
 the alignment in magnetic field was checked and adjusted.
 The fit was then used on blocks of 10 tracks for a fine
 scan of the beam energy delivered by the SPS to the
 Test Beam setup.  
 This was expressed in normal-impact radius of
 curvature equivalent\footnote{The normal-impact radius of
 curvature equivalent is the radius of curvature of the
 particle's track, should the particle impact the field
 orthogonaly. It is a measure of the particle-momentum,
 preferred in the context of multiple tracks, when each track impacts
 the field at a different
 angle.}, and it is shown in figure
  \ref{fig:rrr} (bottom-right).

  \section{Conclusions}

  Analytical methods were shown to have a dramatic impact
 on the speed of
 line and helix fits, bringing down CPU usage by
 3-4 orders of magnitude.
 The  methods developed were successfully tested 
 in the alignment and data reconstruction of the Test Beam
 results for the ATLAS
 Pixel Detector MCM-D modules, and can be of great impact for
  high precision
 alignment and reconstruction in all pixel detectors.
 Such routines could be used for instance by the
 ATLAS Inner
 Detector to specify precision $\vec{p}_\perp$ corrections to the
 $m_{\pi}$ evaluated mass of B-vertices, in order to strongly suppress 
 c-decay backgrounds.

  \section{Acknowledgements}

   I am thankful to the High Energy Physics group
 of the Wuppertal University - in particular to Prof. Dr. K.-H. Becks,
 for
 the kind hospitality and facilities provided
 during completion of this
 work, 
 as well as to
 the ATLAS Pixel Detector Collaboration for the opportunity
 of engaging in the Test Beam activity.
 I would especially like to thank the Alexander
 von Humboldt Foundation for support during this
 period and for
 the opportunity of better knowing German research and
 culture.

%
%
%

%

\end{document}